\documentclass[12pt]{article}

\begin{document}
\renewcommand{\thefootnote}{\fnsymbol{footnote}}

\vspace{2cm}

\begin{center}
{\large\bf SUSY breaking in $S^1/Z_2$ orbifold
models\footnote{Talk at the 9th International
 Conference on Supersymmetry and Unification of Fundamental
 Interactions (SUSY01), Dubna, Russia, June 11-17, 2001}}
\vspace{1cm} \\

M.Chaichian${}^{a}$ \footnote{E-mail:
Masud.Chaichian@helsinki.fi}, A. Kobakhidze${}^{a,c}$ \footnote{E-mail:
Archil.Kobakhidze@helsinki.fi}
and
M. Tsulaia${}^{b,c}$ \footnote{E-mail:
tsulaia@thsun1.jinr.ru}\\
\vspace{0.5cm}

${}^a${\it HEP Division, Department of Physics, University of Helsinki \\
and Helsinki Institute of Physics,}\\
{\it Helsinki, 00014, Finland }\\
\vspace{0.5cm}

${}^b${\it Bogoliubov Laboratory of Theoretical Physics, JINR} \\
{\it Dubna, 141980, Russia}\\
~\\
${}^c${\it The Andronikashvili Institute of Physics, Georgian Academy
of Sciences,}\\
{\it Tbilisi, 380077, Georgia}

\end{center}

\begin{abstract}
We consider the problem of supersymmetry breaking
   in 5 dimensional  $N=1$ supersymmetric models with $S^1/Z_2$
compactification.
\end{abstract}

\vspace{0.5cm}

The remarkable success in the understanding of non-perturbative aspects of
string theories gives a new insights into the particle phenomenology. One of
the phenomenologically most promising approach has been proposed by Ho\v{r}%
ava and Witten within
the 11-dimensional (d=11) supergravity compactified on S$^{1}$%
/Z$_{2}$ orbifold \cite{HW}.
Below we study the problem of   breaking of a rigid supersymmetry in
$d=5$ field-theoretic limit of the Ho\v{r}ava-Witten
compactification.
Consider the $d=5$ off-shell $N=1$
hypermultiplet ${\cal H}=\left( h^{i},\psi
,F^{i}\right) $ which consists of  scalar field $h^{i}$
and auxiliary field $F^{i}$, both being $SU(2)$ doublets and
  Dirac fermion $\psi =\left( \psi
_{L},\psi _{R}\right) ^{T}$.
These fields form two $d=4$ $N=1$  chiral multiplets $H_{1}=\left( h^{1},\psi
_{L},F^{1}\right) $ and $H_{2}=\left( h^{2},\psi _{R},F^{2}\right) $
and  are described by the $d=5$ $N=1$  SUSY Lagrangian:
\begin{equation}
{\cal L}_{hyper}^{(5)}=\left( \partial _{M}h^{i}\right) ^{+}\left( \partial
^{M}h^{i}\right) +i\overline{\psi }\Gamma ^{M}\partial _{M}\psi +\left(
F^{i}\right) ^{+}\left( F^{i}\right).  \label{h2}
\end{equation}
 Let us compactify the fifth
 coordinate $x^4$ on the orbifold $S^1/Z_2$ and define the
transformation properties of fields entering in the hypermultiplet under the
discrete $Z_{2}$ orbifold symmetry,   which acts on the fifth
coordinate as $x^{4}\rightarrow -x^{4}$. A generic bosonic field transforms like
$\varphi (x^{m},x^{4})={\cal P}\varphi (x^{m},-x^{4})$
while the fermionic field transforms as
$\eta (x^{m},x^{4})={\cal P}i\sigma ^{3}\eta (x^{m},-x^{4})$
where ${\cal P}= \pm1$ is an intrinsic parity.
Asserting ${\cal P}= 1$ to the components of multiplet $H_1$
and ${\cal P}= - 1$ to the components of mulitplet $H_2$, one can see that the bulk
Lagrangian is invariant under the action of the parity operator \cite{CKT}.
On the other hand at the orbifold fixed point $x^4=0$ one is left with
N=1 supersymmetry acting on the $d=4$
  chiral multiplet $H_1$
 with the modified auxiliary field $ \tilde F = F^{1}+\partial _{4}h^{2}$.
Analogously, considering the  $d=5$ $N=1$ vector multiplet
which can be decomposed into $d=4$ $N=1$ vector
$V=\left( A^{m},\lambda
_{L}^{1},X^{3}\right) $ and  chiral  $ \tilde \Phi =\left( \Sigma
+iA^{5},\lambda _{L}^{2},X^{1}+iX^{2}\right)$ multiplets and asserting
${\cal P} = 1$ for components of $V$ and ${\cal P}=-1$ for components
of $\tilde \Phi$, the bulk Lagrangian
\begin{equation}
{\cal L}_{Y.M.}^{(5)} =-\frac{1}{2g_{5}^{2}} F^2_{MN}
+\frac{1}{g_{5}^{2}}\left( \left( D_{M}\Sigma \right) ^{2}+
\overline{\lambda }i\Gamma ^{M}D_{M}\lambda  +\left( X^{a}\right)
^{2}- \overline{\lambda }\left[ \Sigma ,\lambda \right]
\right) \label{v3}
\end{equation}
is invariant under the action of the parity operator, while
on the boundary $x^4=0$ we obtain $N=1$ supersymmetry \cite{GAUGE}
realized on the vector multiplet
$V$ with $D=X^{3}-\partial _{4}\Sigma$.

 The mechanism
of supersymmetry breaking in these models \cite{CKT} is based on
a fact that any field configuration that breaks translational
invariance and is not a BPS state breaks supersymmetry totally as
well \cite{DS}. Such
a stable non-BPS configurations with a purely finite gradient energy could
appear in a compact spaces as well (or, more generally, in spaces with
finite volume) if there exist moduli forming a continuous manifold of
supersymmetric states.
 Consider first the case of the pure hypermultiplet in
the  $d=5$ bulk.
Besides the  trivial
vacuum configuration $<h^{2}>=0$
 there can exist  non-trivial
configuration
\begin{equation}
<h^{2}>=\epsilon x^{4},  \label{b1}
\end{equation}
where $\epsilon $ is an arbitrary real constant.
The configuration (\ref{b1}) is odd under the $Z_{2}$ orbifold
transformation  and  breaks translational invariance
in $x^{4}$ direction. However, the configuration (\ref{b1}) does not satisfy
the ordinary periodicity condition on a $S^{1}$ circle,
 but rather the modified one,
\begin{equation}
h^{2}(x^{m},x^{4}+2 \pi R)=h^{2}(x^{m},x^{4})+2\epsilon \pi R ,  \label{b2}
\end{equation}
 The Lagrangian density ${\cal L}_{hyper}^{(5)}$
  remains single-valued and periodic.
Thus, if we assume that $h^{2}$ and its
superpartner $\psi _{R}$ are defined modulo $2\epsilon \pi R$ on $S^{1}/Z_{2}$
 then the configuration (\ref{b1}) will be perfectly
compatible with $S^{1}/Z_{2}$ orbifold symmetries. The configuration
(\ref{b1}) is stable (the variation of corresponding energy functional equals
to zero) and  spontaneously breaks $N=1$ supersymmetry since
$<F^{1}+\partial_{4}h^{2}>=\epsilon \neq 0$.
Analogously there could be also $x^{4}$-independent stable
configuration
\begin{equation}
  <h^{1}>=\epsilon r\sin \theta e^{i\varphi }
  \end{equation}
which breaks  $N=1$ supersymmetry
 on the boundary wall
and is compatible with $S^{1}/Z_{2}$ orbifold symmetries.

 The solution similar to (\ref
{b1}) can be obtained as well for the $Z_{2}$-odd scalar $\Sigma +iA_{5}$ in
the case when vector supermultiplet lives in the bulk and non-trivial
boundary condition analogous to (\ref{b2})  is
assumed.
Note also that the case of interacting gauge
fields in the bulk actually reduces to
the free theory considered above. Indeed
the most general Lagrangian for supersymmetric gauge theories in $d=5$
is expressed
via the holomorphic prepotential being at most cubic \cite{SEIB}
i.e.,
${\cal F}(\Phi )=\frac{4\pi }{g_{5}^{2}}\Phi ^{2}+\frac{c}{3}\Phi ^{3}$.
However the $S^1/Z_2$ orbifold symmetry requires $c=0$ and one is left again with the
free theory described by  (\ref{v3}).

Let us now consider the $N=1$ chiral superfield $\Phi =\left( \phi ,\chi
_{L},F_{\Phi }\right)$ localized on the $d=4$ boundary $x^{4}=0$.
The total Lagrangian has the form
\begin{equation}
{\cal L}_{hyper}^{(5)}+\left[ {\cal L}_{\Phi }^{(4)}+{\cal L}_{\Phi
H_{1}}^{(4)}\right] \delta (x^{4}),
\end{equation}
where ${\cal L}_{\Phi H_{1}}^{(4)}(\Phi ,H_{1})$ describes the interactions
between the chiral superfields $\Phi =\left( \phi ,\chi _{L},F_{\Phi
}\right) $ and $H_{1}=\left( h^{1},\psi _{L},F^{1}+\partial _{4}h^{2}\right)
$ on the boundary $x^{4}=0$ through the superpotential $W_{\Phi H_{1}}(\Phi
,H_{1})$, while ${\cal L}_{\Phi }^{(4)}$ is the usual $d=4$ Lagrangian
for the chiral superfield with the superpotential $W_\Phi$.
Analyzing the equations of motion \cite{CKT}
one can see that
 if $<F^{1}>\neq 0$, and $<F_\Phi>=0$ then the degeneracy in $<h^{2}>$ is
removed and for
\begin{equation}
 <\frac{\partial W_{\Phi H_{1}}}{\partial h^{1}}>=\alpha =const
 \end{equation}
  we get the supersymmetry preserving configuration
\begin{equation}
<h^{2}>=-\alpha \theta (x^{4}).
\end{equation}
Analogously  adding to (\ref{v3}) the FI term
$-2\eta \left( X^{3}-\partial _{4}\Sigma \right) \delta (x^{4})$
one can find supersymmetry preserving configuration
with \begin{equation}
<X^3> = g_5^2\eta \delta(x^4) \quad and \quad
 <\Sigma> = g_5^2\eta \theta(x^4).
 \end{equation}

\vspace{1cm}

\noindent {\bf Acknowledgments.}
The work of M.C. and A.K. was supported by the Academy of Finland
under the Project 163394 and that
of M.T. was supported by RFBR grants \#99-02-18417 and \#01-02-06531
and INTAS grant \#00-00254.


\begin{thebibliography}{99}



\bibitem{HW}  P.~Ho\v{r}ava and E.~Witten,
{\em Nucl. Phys.} {\bf B460} (1996) 506;
{\em Nucl. Phys.} {\bf B475} (1996) 94.

\bibitem{CKT}  M.~Chaichian, A.~Kobakhidze and M.~Tsulaia,
{\em Phys. Lett.} {\bf B505} (2001) 222.

\bibitem{GAUGE}  E. A.~Mirabelli and M.~Peskin,
{\em Phys. Rev.} {\bf D58} (1998) 065002.

\bibitem{DS}  G.~Dvali and M.~Shifman,
{\em Nucl. Phys.} {\bf B504} (1997) 127.

\bibitem{SEIB}  N.~Seiberg,
{\em Phys. Lett.} {\bf B388} (1996) 753.

\end{thebibliography}
\end{document}